\makeatletter
\newcommand{\linebreakand}{%
    \end{@IEEEauthorhalign}%
    \hfill\mbox{}\par
    \mbox{}\hfill\begin{@IEEEauthorhalign}%
}
\makeatother

\documentclass[conference]{IEEEtran}
\IEEEoverridecommandlockouts

\usepackage{listings}
\usepackage{color}
\definecolor{lightgray}{gray}{0.9}

\usepackage{enumitem,kantlipsum} 
\usepackage{placeins}
\usepackage{makecell, multirow}
\usepackage{mwe}
\usepackage{subcaption}
\usepackage{multicol}
\usepackage{booktabs,tabularx}
\usepackage{amsmath,amssymb,amsfonts}
\usepackage{float}
\usepackage[maxbibnames=99]{biblatex}
\usepackage[utf8]{inputenc} 
\usepackage{fancyhdr}
\usepackage{blindtext}
\usepackage{csquotes}
\usepackage[misc]{ifsym}  
\usepackage{braket}
\usepackage[
  separate-uncertainty = true,
  multi-part-units = repeat
]{siunitx}
\usepackage{url}
\usepackage{hyperref}
\usepackage{cleveref }
\usepackage{algorithmic}
\usepackage{algorithm}
\usepackage{graphicx}
\usepackage{textcomp}
\usepackage{xcolor}    
\usepackage{caption}
\usepackage{tikz}
\usepackage{pgfplots, pgfplotstable}
\pgfplotsset{compat=1.17}

\usepackage{tabularx,booktabs}
\newcolumntype{C}{>{\centering\arraybackslash}X} % centered version of "X" type
\usepackage{lipsum}
\def\BibTeX{{\rm B\kern-.05em{\sc i\kern-.025em b}\kern-.08em
    T\kern-.1667em\lower.7ex\hbox{E}\kern-.125emX}}
     
\begin{document}

% \title{Automatic Identification of Social Media Political Articles to Combat Anti-Government Activities}
\title{Towards Identification of Political Posts in Social Media Text Using Deep Learning Models}

\author{%
    \IEEEauthorblockN{Sultan Ahmed}
    \IEEEauthorblockA{%
        \textit{Dept. of Information Systems}\\
        \textit{University of Maryland,}\\
        \textit{Baltimore County}\\
        Maryland, USA\\
        sultan.ahmed@umbc.edu}
    \and
    \IEEEauthorblockN{Salman Rakin}
    \IEEEauthorblockA{%
        \textit{School of Computing}\\
        \textit{Queen's University}\\
        \textit{Kingston}\\
        Ontario, Canada\\
        salman.rakin.kd@queensu.ca}
    \and
    \IEEEauthorblockN{Md. Mostofa Akbar}
    \IEEEauthorblockA{%
        \textit{Department of CSE}\\
        \textit{Bangladesh University}\\
        \textit{of Engineering \& Technology}\\
        Dhaka, Bangladesh\\
        mostofa@cse.buet.ac.bd}
    % \and
    %  \IEEEauthorblockN{Md Osman Gani}
    % \IEEEauthorblockA{%
    %     \textit{Dept. of Information Systems}\\
    %     \textit{University of Maryland,}\\
    %     \textit{Baltimore County}\\
    %     Maryland, USA\\
    %    mogani@umbc.edu}
}
% }
\maketitle

\begin{abstract}
The increasing growth of social media provides us with an instant opportunity to be informed of the opinions of a large number of politically active individuals in real-time. We can get an overall idea of the text nature of these individuals on governmental issues by analyzing the social media texts. Nowadays, different kinds of news websites and popular social media such as Facebook, YouTube, Instagram, etc. are the most popular means of communication for the mass population. So the political perception of the users toward different parties in the country is reflected in the data collected from these social sites. In this work, we have extracted three types of features, such as the stylometric feature, the word-embedding feature, and the TF-IDF feature from the text and applied different kinds of machine learning models including deep learning models. Finally, the deep learning models are compared against traditional machine learning classifiers during identification of political nature from the text. We have compared our methodology with the research work in different languages. Among them, the word embedding feature with LSTM outperforms all other models with 88.28\% accuracy.
\end{abstract}


\section{Introduction}\vspace{0pt}

Text is the most important means of communication in today’s world. Popular online social networking sites such as Facebook, X, LinkedIn, etc. are mainly text-based. The rapid growth of social media has created enough opportunities to share information across time and space. Users are now more comfortable contributing to the content of social media websites and posting their own material.   

With the constant flow of information on the internet, individuals are inclined to consume a greater amount of content from various social media posts and the accompanying comments. 
Nowadays, the people of Bangladesh heavily rely on social media~\cite{tasnim2021political}.   They share a huge amount of political information in these social medias.   Individuals have the ability to share and articulate their opinions or critique political news from their own perspective using a global platform. From individual person text, we can often understand whether the text is political post or neutral post by examining the context of the post, as an individual’s word often reveal nature of the text~\cite{iyyer2014political}. 

The classical approach to feature extraction for opinion mining from textual data is to identify unique stylometric features of written texts. The underlying assumption here is that each author has unique writing styles that are relatively fixed and barely change with
time. So we can use stylometric features to uniquely identify the writing style of the author~\cite{shalabi_kanaanbt}. Along with the stylometric feature, we will use the TF-IDF vectorizer and Word Embedding approach to identify political ideology from textual data.

In this work, we are interested in designing an intelligent model that can analyze and predict whether a given text is political or neutral on the basis of the context understanding from social media platform. The model can take any textual data and predict whether it is a political or neutral post. Based on the prediction, the political ideology of that user can be detected easily. We are interested in addressing the political ideology problem from social media Bangla text. To the best of our knowledge, only one paper has previously addressed this problem. 

\begin{figure*}[!t]
    \centering
    \includegraphics[width=\textwidth]{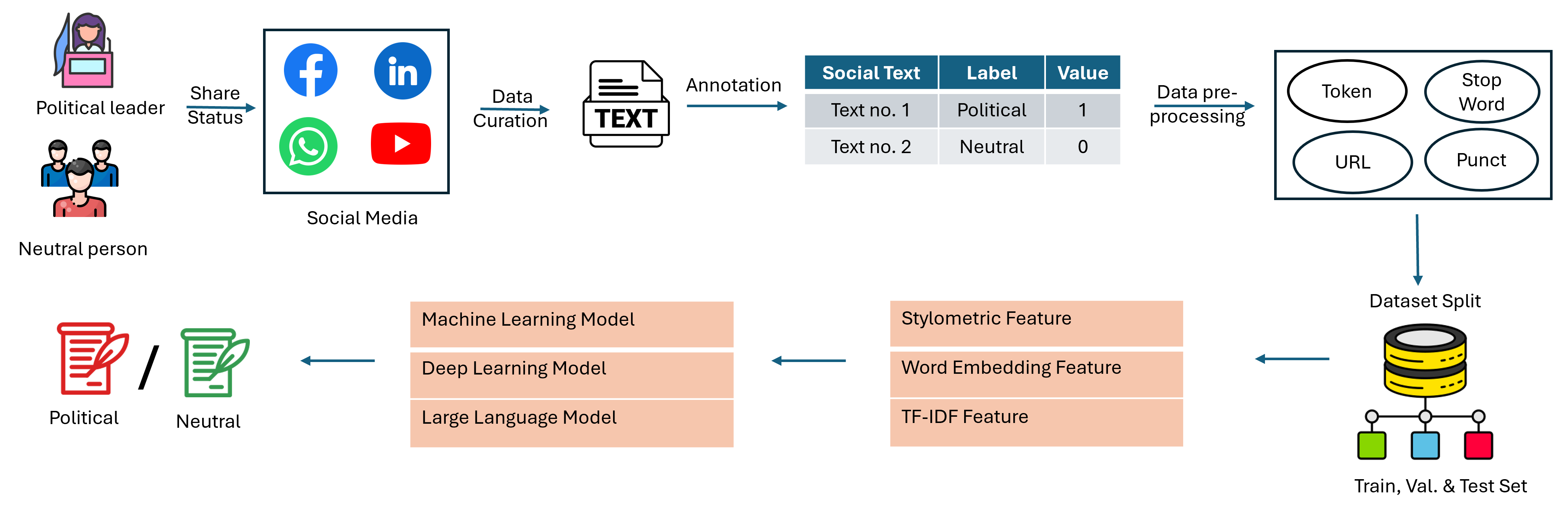}
    \caption{Proposed Methodology for Identification of Political Personality from Social Media Text}
    \label{fig:problem_statement1}
\end{figure*}

% \begin{figure}[H] % 'H' option from float package to control placement
% \centering
% \includegraphics[width=8cm]{Tex_Code/1.intro/problem_statement.png}
% \caption{Problem Statement}
% \label{fig:problem_statement1}
% \end{figure}

Although political post identification has been widely studied in different languages, it is still understudied in the Bangla language. Bangla language is one of the most widely spoken and culturally rich languages. This language is the 7\textsuperscript{th} most spoken language~\cite{spoken_language} of the world and the native language of Bangladesh. However, this is not the only reason to study political post identification problem in the Bangla language. The problem associated with the Bangla language and the relatively under-developed field of Bangla Natural Language Processing (NLP) makes it more challenging to study such problem for Bangla language.

In this work, we will follow the following steps to study the political nature identification problem from social media text. We will create a dataset from Facebook containing political posts and neutral posts. We will then pre-process the data and extract features from the data in 3 ways. One is called the Bag-of-words technique. Another one is computing stylometric features to capture the writing style of the author. The third one is using a word embedding approach to convert text into a feature vector.

In recent years, deep learning-based recurrent neural models have been used to automate political nature of the post due to their performance in building models. These models do not require to be provided with pre-defined handpicked features. Instead, they can learn useful features from the data by themselves~\cite{Bsir_zrigui}.\endgraf 

In this work, we will use deep learning recurrent models to automate political nature determination from Facebook textual data. Specifically, we will use LSTM and GRU models from deep learning recurrent neural network models. From the traditional model, we will use SVM and
NB models. Then the performance of the deep learning model is compared to traditional machine learning models.

The possible contributions to this work are as follows: 

% \begin{itemize}
%   \item We will create a dataset that contains user political text or neutral text in the Bangla language.
%   \item We will design a system that aims to identify political ideology from social media comments or text in the Bangla language. 
%   \item We will propose several types of stylometric features as political post indicators for the Bangla language. We will design a set of measures to infer political ideology from short writings through extensive experiments.
%   \item We will apply machine learning classifiers and deep learning models to identify political bias in short writing. 
% \end{itemize}

\begin{itemize}
  \item We will create a dataset that contains user political text or neutral text in the Bangla language.
  \item We will design a system that aims to identify political nature from social media text in the Bangla language.
  \item We will propose several types of stylometric features as political post indicators for the Bangla language. However, this methodology can be used to identify political nature of the post in any language. 
  
  \item We will apply machine learning classifiers and deep learning models to identify political bias in short writing.

\item  We will design a set of measures to infer political nature from social media text through extensive experiments.

\end{itemize}

The rest of this paper is organized as follows: Section II overviews the related works of the political ideology determination problem. Section III discusses the mathematical background of the problem. In section IV, we have proposed our detailed
solution. Section V presents the experimental results. Finally, in Section VI, we provide project planning and schedule.

% Section V is not completed yet.
% However, in section V, we will conclude our findings with a
% discussion of the obtained observations and the future
% directions of this work.

\section{Related Works}

In ~\cite{wang2013depression}, the authors conducted a depression analysis in Chinese. In their endeavor, Psychological and Machine Learning knowledge were combined. The authors opted for Psychologists who assisted 90 depressive and 90 non-depressed Micro-blog users in collecting a total of 6013 micro-blogs. Their model's precision was 80%. 

Abdul et al. ~\cite{abdul_hasib_arif} proposed using a Long Short-Term Memory Recurrent Neural Network (LSTM-RNN) to analyze Bangla social media posts for depression. They gathered Bangla tweets from Twitter in order to compile the dataset required for this endeavor. This dataset contained 1968 tweets, which was insufficient for a deep learning model to perform adequately. In order to improve the performance of the model with this short dataset, the data were stratified so that one depressive text was followed by one non-depressive text. They experimented with dividing the dataset into 80 percent for training, 10 percent for validation, and 10 percent for model testing. They optimized four hyperparameters of the trained model (LSTM size, batch size, number of epochs, and number of layers) for maximum classification accuracy. The evaluation of the model revealed an accuracy of 86.3\%.  

Authors performed the Gated Recurrent Neural Network algorithm on the same dataset in another work~\cite{uddin2019depression} to predict depressive Bangla text. As in previous work, They worked with the hyper-parameters of the GRU model and achieved approximately 75\% classification for this task. 

Billah et al.~\cite{billah2019depression} collected depressed and non-depressed posts from Facebook manually and applied SGD classifier, Multinomial Naïve Bayes, Logistic Regression, and Linear SVC to detect social media post whether it was depressive or not. During the treatment of patients,  Psychologists prefer some linguistic features that may help to detect depression. For example, depressed people usually use words like “me’, “I”, “myself” etc. which actually represent their self-centered thinking focusing on themselves rather than other people. The authors collected these types of posts to enrich their dataset which consisted of 1000 texts having depressive and not depressive texts. They applied several pre-processing steps to clean the data like punctuation removal, normalization, tokenization, etc. They applied Unigram, Bigram, and Emoticon features in their dataset. They had achieved the highest 77.9\% classification accuracy for the SGDC classifier.

Hassan et al. ~\cite{hassan_Jamil_2017sentiment} developed an automated system to detect depression levels of people from social media posts.  They removed the stop words and applied N-grams, POS tagging, and Negation feature extraction techniques to transform the text into a word vector. Finally, SVM, Naïve Bayes, and Maximum Entropy are applied to the dataset to classify depressive tweets where SVM showed the highest 91\% accuracy. 

Recent studies have also explored the application of Large Language Models (LLMs) to text classification tasks, particularly in zero-shot and few-shot settings. Mu et al.~\cite{mu2024navigating} investigated the effectiveness of LLMs in computational social science classification and found that their zero-shot performance did not consistently match that of fine-tuned transformer-based models, with classification results also being sensitive to prompting strategies. Beyond classification performance, the reliability of generative language models remains another important concern, particularly due to their tendency to produce hallucinated information. Some recent approaches ~\cite{rakin2025leveraging} leverage RAG over LLM and investigated domain adaptation for NLP tasks and demonstrated improvements in reducing hallucinations. These studies highlight the continuing challenges of achieving reliable and effective language understanding across different NLP applications.

Lia et al.~\cite{lia2025read} introduced BanglaBias, a benchmark study of 200 Bangla articles curated from news outlets and blogs for political stance detection. Each article was annotated as government-leaning, government-critical, or neutral. The authors evaluated 28 pretrained language models using prompts that infers political stance and a brief rationale. This study neither trains any language model nor it studies any individuals social media text. 

In this study, we want to identify whether individual Bangla social media posts contain political content. We formulate this task as binary classification and train traditional machine learning classifiers and deep learning models on labeled posts.

\section{Methodology}
\begin{figure}[htp]
    \centering
    \includegraphics[width=\columnwidth]{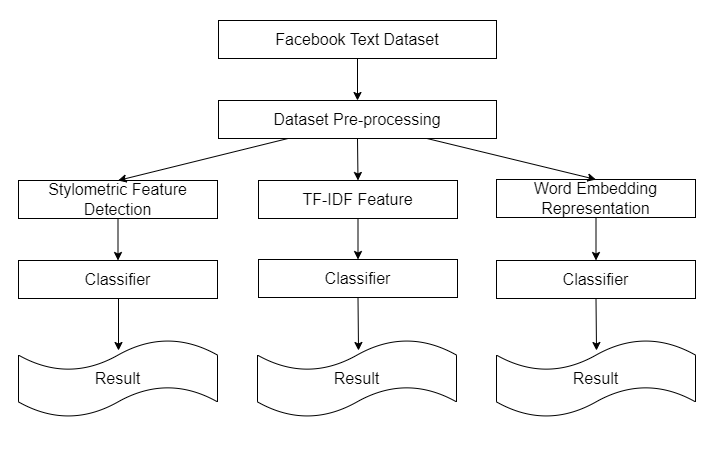}
    \caption{Steps followed in our solution}
    \label{fig:3.methodology}
\end{figure}

This section presents a detailed overview of three feature extraction techniques. One is the Bag-Of-Words feature and another is the Stylometric feature approach and the third one is the word embedding approach. We first label the dataset and then compute the feature after pre-processing the data. The feature is then fed into the machine learning model. Then we provide the architecture of the model.    Fig~\ref{fig:3.methodology} presents an overview of our proposed solution.  

\subsection{Dataset Creation}\label{AA}

Like any other TC problem, the first step in the political hate speech detection problem is to build a large dataset. 
We have collected 1980 texts from this work~\cite{karim2020BengaliNLP} where the number of political class labels is 814 and the number of neutral class labels is 1166 which is listed in Table~\ref{facebook_group_name_table}.\endgraf

\begin{table}[htbp]  
\caption{Data Distributions}
\begin{center}
\begin{tabular}{|p{0.2\textwidth}|p{0.2\textwidth}| }
\hline
 Class Label & Number of posts
 \\ 
 \hline
  Political Hate Post & 
814
 \\ 
 \hline
Neutral Post &
1166
 \\ 
 \hline
\end{tabular}
\label{facebook_group_name_table}
\end{center}
\end{table}

\begin{figure*}[ht!]
\centering
\begin{subfigure}[t]{0.8\textwidth}
\includegraphics[width=\textwidth]{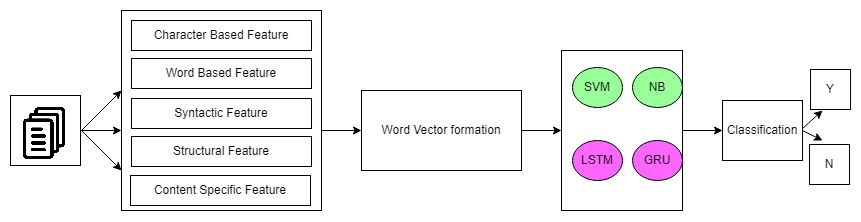}
\caption{Model Architecture for Stylometric Feature} \label{fig:model_architecture_for_stylometric_feature}
\end{subfigure}

\begin{subfigure}[t]{0.8\textwidth}
\includegraphics[width=\textwidth]{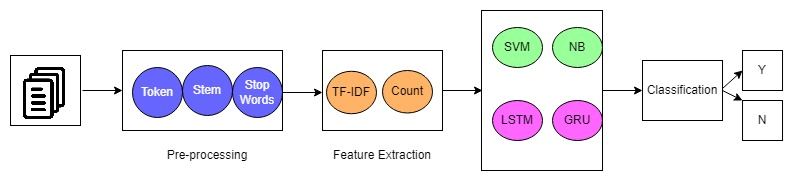}
\caption{Model Architecture for TF-IDF}
\label{fig:model_architecture_for_TF-IDF}
\end{subfigure}

\begin{subfigure}[t]{0.8\textwidth}
\includegraphics[width=\textwidth]{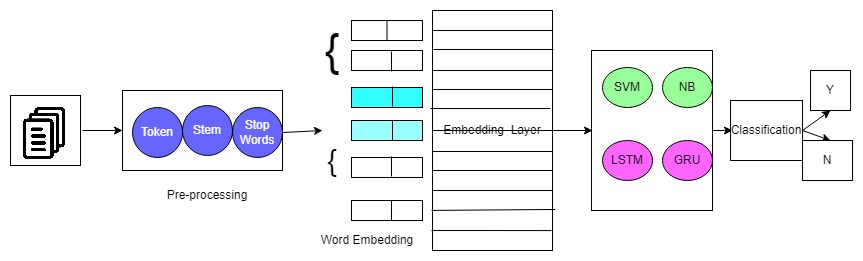}
\caption{Model Architecture for Word Embedding}
\label{fig:model_architecture_for_word_embedding}
\end{subfigure}
\caption{Model Architecture} 
\end{figure*}

\subsection{Pre-processing}
The post obtained from this work~\cite{karim2020BengaliNLP} is noisy and often contains a lot of unnecessary information. So we have applied various pre-processing steps and filtered out all characters except Bangla characters. Then we tokenize our texts and remove stop words from the text. We collect Bangla stop words from GitHub repository~\cite{git_stop_word_bangla} as mentioned in research work~\cite{tripto_ali}. 

Elongated words often contain some context to identify the political inclination of the text. We express our feelings in elongated words. For example, "Greaaat news!!!" has more feelings than "Great news!!!". To maintain the context of the text, we do not apply lemmatization. 

\subsection{Word Vector Formation}
We use deep learning models and traditional machine learning models in our proposed solution. To use these models, we need to convert our text to a word vector. Conversion to a word vector from text is done using stylometric features, TF-IDF features, and word embedding features.

% \begin{enumerate}[wide, labelwidth=!, labelindent=0pt]

% \item   
\subsubsection{\textbf{Stylometric Features Approach:}} 

Stylometric features capture the writing style of both political and neutral posts. We have already computed a large set of stylometric features based on existing works of~\cite{corney_anderson,Otoom_Amer}. These features are categorized into four types: lexical features, structural features, syntactic features, and content-specific features. These four categories of stylometric features have been widely used in research work of ~\cite{abbasi_chen2,abbasi_chen}. 

 \textbf{Lexical features}: Lexical features are the most common set of stylometric features that are intended for stylistics and text readability analysis. These features also signify language assessment and first and second language acquisition. Lexical features consist of word-based and character-based features. These features are concerned with the usage frequency of individual letters, vocabulary richness, entropy measure, consecutive occurrence of words, etc. \endgraf
 
 \textbf{Syntactic features:} These features are primarily intended for identifying writing formation patterns, such as the usage of punctuation marks. These features include the total number of commas, colons, question marks, and exclamations
marks etc.\endgraf

\textbf{Structure-based features:} These features are concerned with how an author organizes the layout of a text; the organization of articles represents different habitual facts of an author, such as paragraph length and use of greetings. Online texts have less content information but richer stylistic information, so these habits are seen to be more prominent in these texts, bearing strong authorial evidence of personal writing styles. \endgraf

\textbf{Content-specific features:} These features represent domain-specific terms. From the study of ~\cite{martindale_mcKenzie}, it has been shown that these features are important in the author's writing pattern formation. For these features, we have first got the feature words as suggested for the Arabic language in ~\cite{shalabi_kanaanbt}. Then we prepared the Bangla feature words by translating these Arabic words using Google Translator service API. We have translated Arabic words into 5 categories: Economy, Policy, Social, Sport, and Negative. 

This translation resulted in many duplicates, flaws, and inconsistencies in the translated lexicons. We have cleared all of these issues by manually inspecting the lexicons.   

\hspace{10pt}For each text of the user, the feature extractor produced a 141 dimension vector to represent the values of the 141 features. As these feature sets contain information on the writing style of a user measured by various methods, the feature values we computed could range from 0 to any positive value.

\hspace{10pt}As we want to ensure all features are treated equally in the classification process, we have normalized the features using the max-min normalization method to ensure all feature values are between 0 and 1:

$$ x^{\prime}_{ij}  = \frac{x_{ij}-min(x_j)}{max(x_j)-min(x_j)} $$

where $x_{ij}$ is the \textit{j}th feature in the \textit{i}th example, min($x_j$) and max($x_j$) are the minimum and maximum feature values of the jth feature separately.

\subsubsection{\textbf{TF-IDF Count Vectorizer Approach}}  TF-IDF (term frequency-inverse document frequency) is an approach that converts the text of the user into a feature vector. We get terms after tokenizing the text of the user. Term frequency represents how many times a specific term occurs in the text. On the other hand, document frequency is the number of documents containing that term.  Term frequency indicates the importance of a specific term in a document. Document frequency indicates how common the term is in~\cite{tf_idf_link}. We have implemented the TF-IDF count vectorizer from the scikit-learn library. We set the number of extracted terms to 1000 to extract features from any text. Suppose we have some text from users. To convert these texts into feature vectors, we have to first identify unique words and count how many times these words occur in each text. Then we have to compute inverse document frequency (IDF) using the following formula. 

$$ idf_i = log\frac{n}{df_i} $$

where $df_i$ represents how many documents contain the term i and n is the total number of documents. Now we will multiply the TF matrix with IDF score to get the vectorized form of each text. All texts are converted into feature vectors using this TF-IDF approach. These vectors can be fed into any machine-learning algorithm.

% \item   
\subsubsection{\textbf{Word Embedding Representation Approach}} Traditionally, in text classification, the bag-of-words (BOW) model is used to extract features from the text. At the time of extracting features from text, the BOW model does not consider grammar, context, or even word order. This model only keeps track of the multiplicity of tokens in text. Authors in~\cite{shalabi_kanaanbt} applied the BOW model with a set of hand-crafted rules to prepare the feature set.  \endgraf
    
However, with the advancement of deep learning models in text classification, the word embedding approach has evolved to capture syntactic and semantic regularities. In this approach, individual words are represented as real-valued vectors in a predefined vector space. In the following, we explain: 1) the methods of the word2vec algorithm, and 2) how the word2vec algorithm can generate word embedding. Finally, we discuss the classification approach after getting word embedding from user text.

\begin{enumerate}[wide, labelwidth=!, labelindent=0pt]
 \item Word2vec:
 
 Word2vec, an efficient algorithm proposed by Google~\cite{mikolov_kai},  can learn a standalone word embedding from a text corpus efficiently maintaining the contextual meaning of words. Word2vec has two model architectures to produce an embedding representation of words. One is Continuous Bag of Words (CBOW), and another is Skip Gram (SG). CBOW Model takes the context of each word as the input and tries to predict the word corresponding to the context. SG predicts the surrounding window of context words based on the current single word. The word vector prediction is not influenced by the order of the context words.
  
\item Word Embedding using Word2vec: \endgraf  
The Word2vec model can capture a lot of information maintaining semantic, conceptual, and contextual relations. We have learned the embedding vector of each word from a user post on Facebook of our dataset using the CBOW and Skip-Gram model. \endgraf

For example, let us have two sentences in our dataset. [I have a book, I love to eat mango]. So first we split the sentence and generated a two-dimensional vector. The two-dimensional vector will be [ [I, have, a, book], [I, love, to, eat, mango]]. Then we pass this two-dimensional vector to the word2vec model. Skip-Gram and CBOW models generate the word vector from this two-dimensional dataset using a window size of 5. The size of the word vector is 300.   We have used the Gensim package to implement the  Word2vec model. This model returns a 300-dimensional vector for each of these words: I, have, a, book, love, to, eat, and mango. We save these word embeddings and later use these embeddings as the weight of the embedding layer. \endgraf
     
\item Classification:

For classifying a text, we first encode each word of the text using a unique number. Then we multiply this encoding vector with the embeddings of words present in s to form the hidden representation of s. These sentence representations are used to train a linear classifier. Specifically, we use the softmax
function to compute the probability distribution over the classes in C. 

\end{enumerate}

\begin{table*}[t]
 \caption{Performance measure for Stylometric Feature: 
}
\label{performance-table-stylometric-feature}
\begin{tabularx}{\textwidth}{@{}l*{2}{C}c@{}} 
\toprule
Model & Feature Vector & Accuracy & F1-score \\
\midrule
LSTM & Styometric Feature & 73.18\% & 61.96\% \\ 
GRU & Styometric Feature & 72.67\% & 64.21\%  \\ 
SVM & Styometric Feature & 78.88\% & 71.35\% \\ 
NB & Styometric Feature & 61.46\% & 37.52\% \\ 
\bottomrule
\end{tabularx} 
\end{table*}

\begin{table*}[t]
 \caption{Performance measure for Word Embedding Feature: 
}
\label{performance-table-embedding-feature}
\begin{tabularx}{\textwidth}{@{}l*{2}{C}c@{}} 
\toprule
Model & Feature Vector & Accuracy & F1-score \\
\midrule
LSTM & Word Embedding Feature & 88.28\% & 85.41\% \\ 
GRU & Word Embedding Feature & 88.23\% & 85.17\%  \\ 
SVM & Word Embedding Feature & 72.72\% & 62.75\% \\ 
NB & Word Embedding Feature & 65.40\% & 16.96\% \\ 
\bottomrule
\end{tabularx} 
\end{table*}

\begin{table*}[t]
\caption{Performance measure for TF-IDF feature:
}
\label{performance-table-tfidf-feature}
\begin{tabularx}{\textwidth}{@{}l*{2}{C}c@{}} 
\toprule
Model & Feature Vector & Accuracy & F1-score \\
\midrule
LSTM & TF-IDF Feature & 69.69\% & 42.30\% \\ 
GRU & TF-IDF Feature & 69.94\% & 46.15\%  \\ 
SVM & TF-IDF Feature & 66.16\% &  37.38\% \\ 
NB & TF-IDF Feature & 67.42\% & 43.67\% \\ 
\bottomrule
\end{tabularx} 
\end{table*}

\begin{table*}
\setcellgapes{3pt}
\makegapedcells
\caption{Comparison of Proposed Method with Previous Works}
\label{table:comparison_of_proposed_method_with_previous_work}
\begin{tabularx}{\textwidth}{ | *{5}{C|} }
  \hline
  \textbf{Language} &  \textbf{Proposed by} &  \textbf{Methodology} &  \textbf{Feature Extraction} & \textbf{Accuracy} \\ 
  \hline 
  Bangla & ~\cite{tasnim2021political}  & None
 &  CBOW, Skip-Gram & 76.22\%   \\
    \hline
   Bangla & ~\cite{rahman2018datasets} & KNN, SVM, RF	
& TF-IDF
& 71.21\%  \\
    \hline
English & ~\cite{baly2020we}  & LSTM, BERT 
 & Word Embedding  & 72.37\%  \\
    \hline
Bangla &  Proposed Method & LSTM, GRU, SVM, NB & Stylometric Feature, TF-IDF Feature, Word Embedding Feature &  88.28\%\\
    \hline
  
 \end{tabularx}
\end{table*}

\subsection{Model Architecture}
We implement two different approaches to identify political ideology from the text. The first one is a deep learning model, and the other is a traditional machine learning model classifier. In the deep learning model architecture, we implement LSTM and GRU models. In the traditional machine learning model, we implement SVM and NB models.

\begin{enumerate}
  \item Model with Stylometric Features: \\
We have prepared word vectors based on the stylometric features. We pass these vectors to the LSTM layer, which has 300 nodes. The output of the LSTM layer is passed to a dense layer. Softmax~\cite{sharma_ochin} is used as an activation function. The optimizer is RMSprop, and binary cross entropy ~\cite{shipra_saxena} is used as a loss function. The same process is repeated for Gated Recurrent Unit(GRU), SVM, and NB models. We have noted down each model's accuracy, F1-score. Fig. \ref{fig:model_architecture_for_stylometric_feature} shows the architecture of the LSTM, GRU, SVM, and NB model with the stylometric feature.

  \item Model with TF-IDF feature:\\ 
We have extracted feature vector from text using the Term Frequency Inverse Document Frequency (TF-IDF) approach to feed into deep learning and traditional machine learning models. We have initialized the TF-IDF vectorizer using n = 2 grams and maximum vocabulary is set to 1000. The user text is converted to feature vectors using this TF-IDF approach. We have shown the architecture of traditional machine learning and deep learning
models with TF-IDF features in figure \ref{fig:model_architecture_for_TF-IDF}.

   \item Model with Word Embedding feature:\\ 

In this architecture, we have first tokenized the user text and taken only the first 100 words. Shorter text is padded with zeros. At the same time, the word embedding of the top 1000 vocabulary words is generated using the word2vec algorithm. These embeddings are set as the weight of the embedding layer. The tokens are converted into sequences and fed into LSTM and GRU models. The output of these models is passed to a dense layer, which is used to detect political ideology from the text. Softmax~\cite{sharma_ochin} is used in the dense layer as an activation function. The optimizer is RMSprop and binary cross entropy~\cite{shipra_saxena} is used as a loss function. The same process is repeated for all the remaining deep-learning models. Figure~\ref{fig:model_architecture_for_word_embedding} shows the architecture of our models with the word embedding features.
 
\end{enumerate}

\section{Experimental Evaluation}
In this section, we have evaluated the performance of our proposed methods for depression detection on Facebook data sets. We compare the performance of deep learning algorithms with traditional machine learning algorithms like Support Vector Machine(SVM) and Naive Bayes(NB). 

\subsection{Experimental Setup}
We have employed Python Keras framework with Tensorflow as a framework to implement deep learning models for training, tuning, and testing. We have also used Gensim package for word2vec model implementation. Scikit-learn package is used to implement traditional machine learning algorithms. Experimental evaluation was conducted on a machine with an Intel Core i7 processor with 1.8GHz clock speed and 8GB RAM. The machine has also an Nvidia GeForce MX150 with 2GB memory and therefore Tensorflow based experiments have fully utilized GPU instructions. Considerable speed can be achieved in Tensorflow based experiments by adding a GPU as shown in ~\cite{JMLR_speed_gpu}.

\subsection{Performance Evaluation and Parameter Tuning}
We have studied the efficiency and scalability of our proposed methods by varying model architectures and feature set vectors. We have measured the performance of the recurrent neural network, LSTM model by generating word vectors using the stylometric feature method.

\subsection{Result Analysis}
\begin{enumerate}
\item Performance of stylometric feature: 

We present the performance of stylometric features along with traditional machine learning and deep learning algorithms in Table~\ref{performance-table-stylometric-feature}. From Table~\ref{performance-table-stylometric-feature}, we can see that traditional machine learning algorithms such as SVM performs better than deep learning algorithm like LSTM. We have used stylometric features that can capture the writing style of different authors.  SVM outperforms all the models because the SVM model with stylometric features can predict political ideology by creating a decision boundary.  Again, LSTM having feedback connections can process the entire sequence of data. Thus LSTM with stylometric features outperforms traditional machine learning algorithms like SVM and NB.

\item Performance of word embedding feature:
The performance of word embedding using the word2vec algorithm is shown in Table~\ref{performance-table-embedding-feature}. From this table, we can see that LSTM and GRU models outperform SVM and NB models. The LSTM model can learn sequential information from feature vectors. Again, word embedding vectorizers can capture the semantic relationships of words. Thus, LSTM with the word embedding approach outperforms all other models. 
  
\item Performance of TF-IDF feature:

We have shown the performance of TF-IDF feature with deep learning and traditional machine learning models in Table~\ref{performance-table-tfidf-feature}. Here, the GRU model outperforms the LSTM, SVM, GRU, and NB models. TF-IDF feature assigns weight to a word based on the number of times it appears in the text. GRU model can capture long-term dependencies in sequential data. So, GRU model with the TF-IDF feature outperforms other models. The highest accuracy is 69.94\% obtained by the GRU model with the TF-IDF feature. 

\item Performance among different models: The performance of different machine learning models associated with features is presented in Figure~\ref{fig:performance_compare_among_different_models}. From Figure~\ref{fig:performance_compare_among_different_models}, we can see that the LSTM model outperforms traditional machine learning models such as SVM \& NB. LSTM with the word embedding feature outperforms all other models. This is expected, as LSTMs can preserve the previous state. SVM performs slightly better than NB in terms of accuracy since SVM works based on the computation of hyperplane equations, which separates data into classes perfectly~\cite{Kusumawati_2019}. The highest accuracy and F1-score obtained by our proposed methods are 88.28\% and 85.41\% respectively.

% \begin{figure*}[ht!]
% \centering
\begin{figure}[t]
\centering
\includegraphics[width=0.5\textwidth]{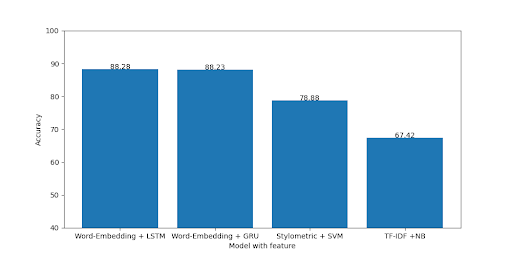}
\caption{Performance compare among different models} \label{fig:performance_compare_among_different_models}
\end{figure}
% \end{figure*} 
 
\item Comparison of Proposed Method with Previous Works:

We have presented a comparison of our methods with other works on determining political ideology from the text in Table ~\ref{table:comparison_of_proposed_method_with_previous_work}. As the number of previous works done on political post detection from social media text in the Bangla language is very low, among ~\cite{tasnim2021political}, ~\cite{rahman2018datasets}, ~\cite{baly2020we} and our proposed methodology, which works are done for Bangla language, our work is showing comparatively better performance.

\end{enumerate}

% \section{Discussion}

% We propose three types of features and deep learning and traditional machine learning models to separate political posts from Social Media Texts. Among them, the LSTM model with the word embedding feature outperforms all other models. The best accuracy obtained by LSTM with the word embedding feature is 88.28\%. Stylometric features tailored to find the writing style of authors are deduced in this work for the Bangla language. We have also shown superior accuracy of the word embedding approach among all other features. However, the approach to finding the research questions of determining political ideology from text acknowledges several limitations and ethical considerations. The methods used in this work may not directly apply to all languages or scenarios. Again, we do not consider emoticons in identifying political ideology from the text. In this work, we put our effort only into determining the political ideology. In future, we can work on finding political party of our country. We can also predict which party is better for the country based on their ideology and citizens opinions towards them. Transformer based machine learning models can be applied to check if better accuracy can be found. 

\section{Conclusion \& Future Work}

Research on identifying political nature of text is ongoing in various languages. However, there is a lack of research in the Bangla language. Again, there is no publicly available dataset to systematically value the political nature of text in Bangla language. We have made our dataset public so that future work in this domain can be done using this dataset. In this study, we have carried our research by applying the stylometric feature, the word embedding feature, and the TF-IDF feature after pre-processing the dataset. Various types of machine learning classifiers and deep learning models are used to identify the political nature of the text. In future, we will extend this methodology to identify political  ideology from text so that forensic department or other security forces can get a preliminary idea about that person. Political party will also get an idea whether the person is a spy or their followers based on the activists textual data. Again, we can predict the political party affiliation from text using these methods. We will also increase the size of the dataset and implement more models so that better accuracy can be obtained.

% \section*{Acknowledgment}
% \input{Tex_Code/6.acknowledge/acknowledgement}    
\printbibliography

\vspace{12pt}
\end{document}